# Learning Geometric Transformations for Parametric Design: An Augmented Reality (AR)-Powered Approach

Zohreh Shaghaghian[1], Heather Burte[2], Dezhen Song[3], and Wei Yan[1]

[1]Texas A&M University, Department of Architecture, College Station, TX 77840, USA
[2]Texas A&M University, Department of Psychological & Brain Sciences
[3]Texas A&M University, Department of Computer Science and Engineering
```
shaghaghian.zohreh@gmail.com
            wyan@tamu.edu
```

**Abstract.** Despite the remarkable development of parametric modeling methods for architectural design, a significant problem still exists, which is the lack of knowledge and skill regarding the professional implementation of parametric design in architectural modeling. Considering the numerous advantages of digital/parametric modeling in rapid prototyping and simulation most instructors encourage students to use digital modeling even from the early stages of design; however, an appropriate context to learn the basics of digital design thinking is rarely provided in architectural pedagogy. This paper presents an educational tool, specifically an Augmented Reality (AR) intervention, to help students understand the fundamental concepts of parametric modeling before diving into complex parametric modeling platforms. The goal of the AR intervention is to illustrate geometric transformation and the associated math functions so that students learn the mathematical logic behind the algorithmic thinking of parametric modeling. We have developed BRICKxAR_T, an educational AR prototype, that intends to help students learn geometric transformations in an immersive spatial AR environment. A LEGO set is used within the AR intervention as a physical manipulative to support physical interaction and improve spatial skill through body gesture.

**Keywords:** Geometric Transformation, Mathematics, Education, Augmented Reality, Parametric Modeling

## 1 Introduction

The superficial knowledge that architectural students have in "digital design thinking" can cause them to adopt excessive trial-and-error approaches when using parametric modeling software, which leads to them not taking full advantages of what the software has to offer [1]. Most students are exposed to parametric modeling software without having the fundamental knowledge of computational design concepts including variables (as the properties of the geometry), parameters (as the members of the function family), and functions (as the math operations). Understanding geometric transformation as one of the essential components in parametric modeling and learning the associated math functions would allow students to improve their knowledge of the mathematical logic behind digital design modeling and utilize parametric modeling software more efficiently and professionally [2]. However, studies have confirmed the difficulty of learning geometric transformations and the related math concepts through common traditional methods [3][4]. Despite the close relationship between spatial thinking and math problem solving, much of mathematics is still taught in a number sense - a collection of mathematical rules and procedures - in many educational systems, and geometry courses often focus on shape attributes rather than spatial reasoning [5]. Spatial methods can be useful for solving mathematical problems when diagrams, drawings, graphs and conceptualization are applicable [5]. Based on APOS (Action-Process-Objects-Schema) theory [6], students with only an action conception in math knowledge are more likely to use a trial-and-error technique to find function outputs instead of conducting an analysis and in-advance prediction to solve a problem. Such superficial techniques could cause architectural students to face difficulty in properly analyzing and understanding the logical algorithmic process behind sophisticated parametric geometry modeling. Modeling software may not provide a competent context to educate the fundamentals, specifically due to the extraneous cognitive load that GUIs (Graphical User Interfaces) impose on learners [7]. On the other hand, physical model interaction has shown significant impact on spatial visualization and reduction of extraneous cognitive load [8][9]. Hence, due to the close relationship between spatial reasoning and mathematics, physical interaction could affect learning math concepts positively.

Augmented Reality (AR) as a mediator tool with the ability to superimpose abstract information over the physical environment provides a spatial intervention that could support embodied learning and virtual augmentation. We have developed an AR educational prototype for teaching the fundamentals of parametric modeling including geometric transformations and the associated math functions using graphical elements (e.g., arrows, tags, highlighting). The prototype, named as BRICKxAR_T is developed on top of the prior work - an AR instruction tool for LEGO assembly [10]. We have developed BRICKxAR_T as a Rotation/Translation/Scale (RTS) puzzle game for teaching geometric transformations and the associated math functions within the constructionist learning environment. A tangible LEGO model is employed as a physical manipulative to support physical interactions.

## 2   Background

### 2.1   Physical interaction in education

In various areas of STEAM (Science, Technology, Engineering, Arts/Architecture, and Mathematics), the effect of physical models on cognitive learning and spatial abilities has been explored. The studies suggest that physical activities increase the creativity of students in design ideation [8][11], reduce the extraneous cognitive load in the creative design process [12], promote students' spatial skill in understanding scale relations between geometries [13], and improve interaction and communication in a collaborative working environment [14]. Physical interaction supports embodied spatial awareness and helps students in mental visualization skills [15]. Psychological studies argue that physical interaction facilitates the epistemic behavior of students, enables them to form embodied abstract metaphors to internalize the data, and helps them strengthen memory retrieval [16]. Multiple research projects have studied the advantage that physical models have over textbooks or computer-based 3D models [9][8], revealing the impact of physical model/interaction on spatial cognition and understanding of complex spatial relations. The results of a study exploring the impact of Tangible User Interface (TUI) versus Graphical User Interface (GUI) found that TUI provides more epistemic actions for designers, promotes spatial cognition, and stimulates design creativity, while GUI restricts designers to only following the design briefs and causes less exploration and discovery between design and solution spaces [8].

### 2.2 Digital and parametric modeling in architectural education

Digital modeling is an essential tool for effective design generation and design exploration. Digital modeling brings a new paradigm of design thinking named "digital design thinking." Parametric design and its ability to generate multiple design alternatives through anchoring design parameters is one of the most innovative achievements of digital modeling. However, during the digital design thinking process, students need to internalize an understanding of parametric design elements (variables, parameters, and math functions) and algorithmic analytical thinking in order to successfully take the full advantage of a parametric design process. Although the current visual programming platforms in parametric modeling may not require in-depth knowledge of computer language syntax, they urge for a proper understanding of input data (design variables), data structure, function components (parameters and equations), and output data for an efficient process and valid results. The importance of spatial transformations and their mathematics - linear algebra including vectors and matrices - for design professionals who wish to utilize and develop efficient computational techniques has been emphasized clearly in "Essential Mathematics for Computational Design" by Rajaa Issa, 2010 [2], a textbook used widely in courses of parametric modeling and in learning of the modeling tools Rhino/Grasshopper.

### 2.3 AR in education

AR applications have been studied in many educational fields, such as physics, mathematics, chemistry, and biomedical sciences, to enhance learning, especially in the situations where students cannot feasibly achieve real-world experiences [17][18]. In these experiments AR is used as an instrument to embody the interpretation of artifacts or abstractions in the physical world to ease the learning process. The intrinsic capability of AR to automatically align the perspective view with the user's relative position provides a context for viewing the digital model from any arbitrary, natural perspective without mentally translating 2D to 3D [19]. Hence, the automatic natural perspective alignment could reduce the extraneous cognitive load that may impose through interpreting 3D models from 2D desktops. A couple of studies have investigated the applications of AR in learning geometry and related mathematics and demonstrated positive impact of AR intervention in geometry perception [20]–[22]. The results of the study done by Dünser et al. (2006) showed insignificant, yet positive, impacts from AR intervention on learning geometry and math [20]. Part of the reason for the insignificant impact might be due to user's interaction with the digital model, which was realized through pencils and panels instead of direct interaction. GeoGebra AR [22] is another AR educational tool for learning geometry and algebra based on the widely used GeoGebra application. GeoGebra AR has a limitation of anchoring virtual geometric models to only physical surfaces instead of arbitrary 3D physical objects (in contrast with BRICKxAR_T that enables physical-virtual model registration, tracking, and interplay). Studies show that AR can enhance teaching quality and contribute to architectural design [23][24], improve the understanding of building science through immersive visualization of building information [25]–[27], advance the understanding of design sustainability and performance-based design [28]–[30], and encourage collaboration and engagement among Architecture/Engineering/Construction (AEC) students [31] [26]. The overall results from studies in literature confirm that students' learning effectiveness improves when the related information is situated spatially and temporally close to the real-world experiment.

Most studies in literature have investigated AR's impact in the architectural industry, design ideation, collaboration, and communication. To the best of authors' knowledge, no research has studied the educational impact of an AR intervention in improving the learning of geometric transformations and allied math concepts in order to enhance architecture students' understanding of parametric and digital thinking.

## 3     Methodology

In this paper, we present an educational AR application to augment abstract, mathematical information found in geometric transformations into a physical environment. The prototype is developed to support three levels of 'motion, mapping and function' in order to create a "network of learning path" for progressive learning of transformation conceptions [32]:

- Motions: AR supports physical actions (using LEGO model as a physical manipulative) to enable embodied learning through a spatial learning experiment.
- Mappings: AR supports visualizing the transformation mapping process through demonstration of the transformation image (physical model) and pre-images (virtual models) with input and output variables (points, lines and surfaces), and transformation parameters (rotation axis, distance, and angle) through displaying of arcs, lines, and notions.
- Functions: AR supports augmentation of mathematical functions of transformation matrices and their multiplications through spatiotemporal alignment of information and physical actions to enhance students understanding of geometric transformations as mathematical functions.

Figure 1 describes the framework of the developed prototype. The AR App is developed for iOS device using the Unity game engine, Apple's ARKit, C# programming, and

Xcode. The last section of the framework, i.e. Deploying, is in the process and will be added in a future study.

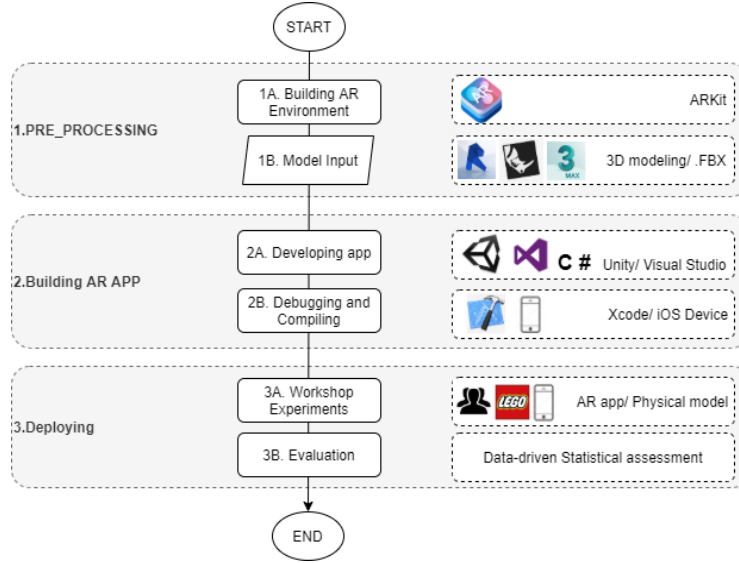

**Fig. 1.** Proposed workflow for the AR app development

### 3.1 Prototype

We have developed a primary prototype for an RTS (Rotation, Translation, and Scale) puzzle game in an AR environment. The prototype is presented in Research Symposium 2020, College of Architecture, Texas A&M, and is publicly visible at YouTube [33]. In this prototype, the three levels of geometric transformations, i.e. "motion, mapping, and function," are realized by visualizing pre-images and images of transformations in the AR environment and displaying the associated math functions that are dynamically aligned with transformations conducted in the integrated physical and AR environment. Graphical information such as dimension lines, rotation arcs, coordinate systems, and notations are displayed in the AR environment in real-time to assist users in visualizing math concepts and tracking the transformations matched with the users' perspective in the physical environment. Two rows of matrices displayed in AR demonstrate the math functions associated with the geometric transformations conducted in the physical and AR environment. The matrices are the common 4×4 transformation matrices that could represent all types of linear transformations such as rotation, translation, scale, reflection, and shear [5], as well as perspective transformation. Figure 2 shows the transformation matrices used in this prototype, including translation, rotation, and scale. The blue highlighted region of the 4×4 transformation matrix (column 1-3 and row 1-3) would be filled by any of the 3×3 matrices of rotation/scale, while the translation vector will fill the orange region (column 4 and row 1-3). The result of the combination of transformations will be conducted through matrix multiplication accordingly.

**Rotation around x_axis**  **Rotation around y_axis**  **Rotation around z_axis**  **Scale 3D**

$$R_x = \begin{bmatrix} 1 & 0 & 0 \\ 0 & \cos(t) & -\sin(t) \\ 0 & \sin(t) & \cos(t) \end{bmatrix} \quad R_y = \begin{bmatrix} \cos(t) & 0 & \sin(t) \\ 0 & 1 & 0 \\ -\sin(t) & 0 & \cos(t) \end{bmatrix} \quad R_z = \begin{bmatrix} \cos(t) & -\sin(t) & 0 \\ \sin(t) & \cos(t) & 0 \\ 0 & 0 & 1 \end{bmatrix} \quad S = \begin{bmatrix} scale & 0 & 0 \\ 0 & scale & 0 \\ 0 & 0 & scale \end{bmatrix}$$

**Translation**

$$V = \begin{bmatrix} x \\ y \\ z \end{bmatrix}$$

**4×4 transformation matrix:**

|  | col(1) | col(2) | col(3) | col(4) |
|---|---|---|---|---|
| row(1) | + | + | + | x |
| row(2) | + | + | + | y |
| row(3) | + | + | + | z |
| row(4) | + | + | + | + |

**Fig. 2.** Transformation matrices

In BRICKxAR_T, the first row of matrices shows a transformation matrix and associated mathematical functions (i.e. linear algebraic operation of matrix multiplication) applied on the physical model in the physical environment in real-time. The second row corresponds to the transformations of the digital model in the AR environment. In the beginning, both matrices are identity matrices, while a digital wireframe model superimposes the physical LEGO model using marker registration (Figure 3. Left). When the user starts to play with the physical model and apply actions on the model, for example, move or rotate the model, the wireframe model stays in the primary position representing the pre-image of the transformation. The tracking graphics (i.e., dimension line, rotation arc, and notations) appear to represent the mapping relation between the pre-image (wireframe model) and the image (physical model). At the same time, the user can observe the mathematical functions of the applied transformations where numbers match the notations displayed on dimension line and rotation arc (Figure 3. Right) for developing an intuitive understanding of the transformations. This approach generates a meaningful context for abstract mathematical functions aligned with the studies that reveal the benefits of contextualization in presenting mathematics content [34][35].

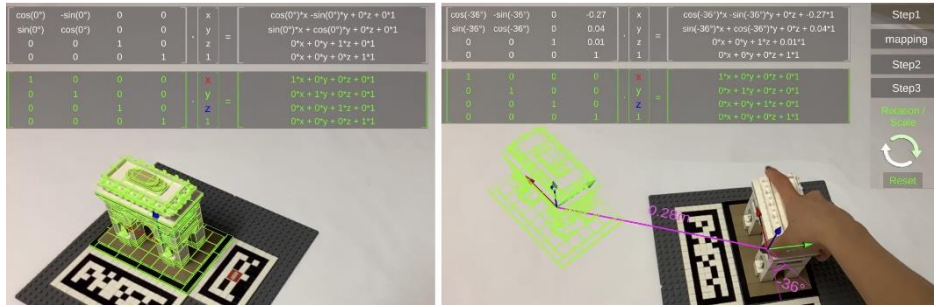

**Fig. 3.** Left: The wireframe model superimposed on the LEGO model using marker registration. Right: Transforming (translating/rotating) the physical model freely with hand.

Later, students can apply transformations on the virtual model for matching the transformations of the physical model through interacting with the members of the function family using sliders: for example, touching x, y, and z of the translation vector or tapping the transformation logo to apply rotation and scale through modifying the values of the selected parameters. Figure 4 shows the parameters corresponding to the translation operation. The colors of the translation vector parameters match with the associated coordinate system to visually help students in matching the math information to the action.

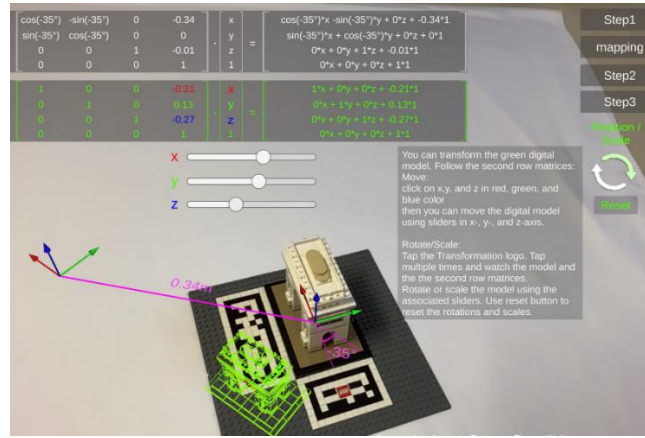

**Fig. 4.** Interacting with parameters of the matrix function to transform the virtual model.

To apply other transformations, i.e., rotation and scale, students can tap on the transformation logo displayed on the screen. Each time the logo is tapped, the associated transformation matrix and the corresponding slider to change the selected parameter (i.e., rotation angle or scale factor) appear on the screen. For rotation, the corresponding rotation axis and plane are graphically highlighted as well as the color of the logo that get matched with the corresponding rotation axis. These graphical representations intend to catch students' visual attention. Figure 5 shows the graphical information and associated matrices for rotation around x- and y-axis, and Figure 6 shows the same information for rotation around z-axis and 3D scale. The three coordinate systems in the figures belong to the physical model (image), the original position of the model (pre-image) and the digital model respectively. The last two coordinates are overlapped in figures 5 to 7.

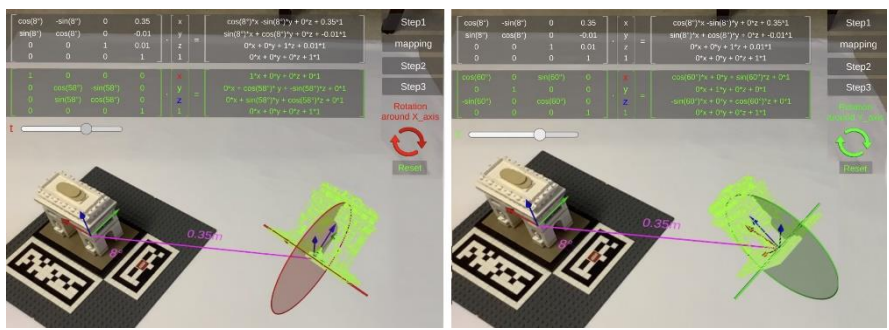

**Fig. 5.** Left: Rotation around x-axis; Right: Rotation around y-axis

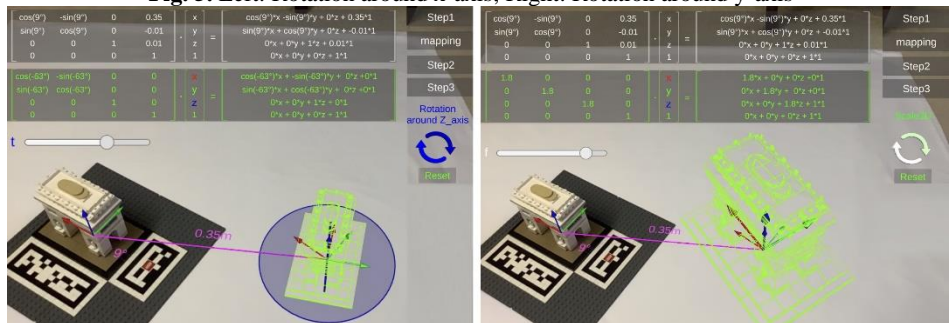

**Fig. 6.** Left: Rotation around z-axis; Right: 3D scale

In addition, the spatial mapping concept can be visualized using a set of representative points on the pre-image and their corresponding mapped points on the image (Figure 7).

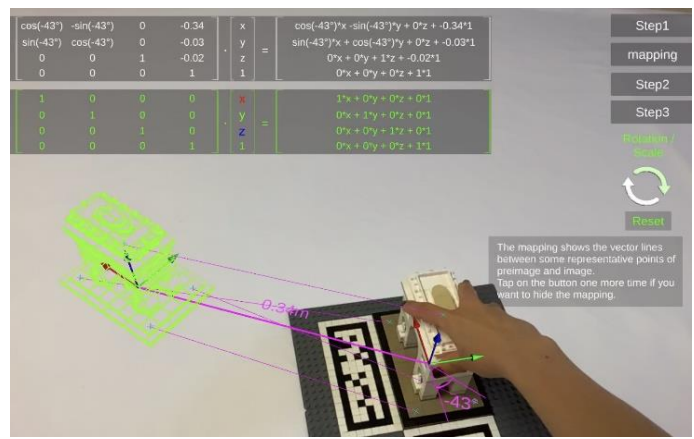

**Fig. 7.** Visualize mapping operation between pre-image and image using mapped points on the models or in the space.

In this phase, the students can play with multiple parameters such as translation axes (x, y, and z), rotation axes (x, y, and z), and the scale factor, and trace the associated math matrices corresponding to the wireframe model, i.e., the second row of matrices in green color. The task assigned to this part is to practice the AR registration (physical and virtual model alignment) method that is normally done automatically by the AR technology using the camera and motion sensors on the AR device. Students need to modify function parameters to implement the correct transformations so that the two models (virtual and physical) align again. In order to exactly match the two models, students need to compose the transformation matrix of the virtual model (second row) so it matches with the transformation matrix of the physical model (first row) in the transformed position. The original position of the LEGO model (the primary pre-image) is displayed with a world coordinate system in this prototype where the model is registered in the beginning. Figure 8 shows from a third-person view a student playing with BRICKxAR_T using an iPad with AR enabled.

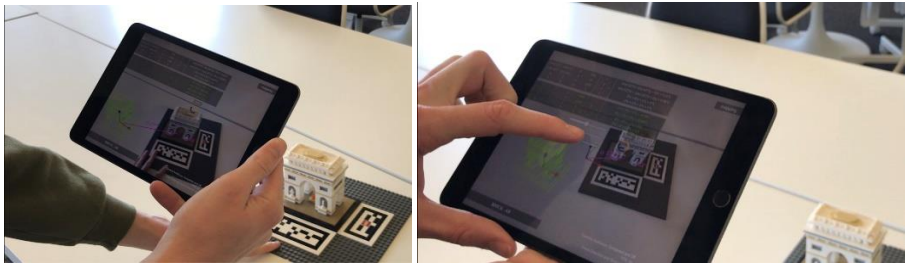

**Fig. 8.** Student playing with BRICKxAR_T through an iPad

Figure 9 shows screen shots of the App where the student is practicing AR registration.

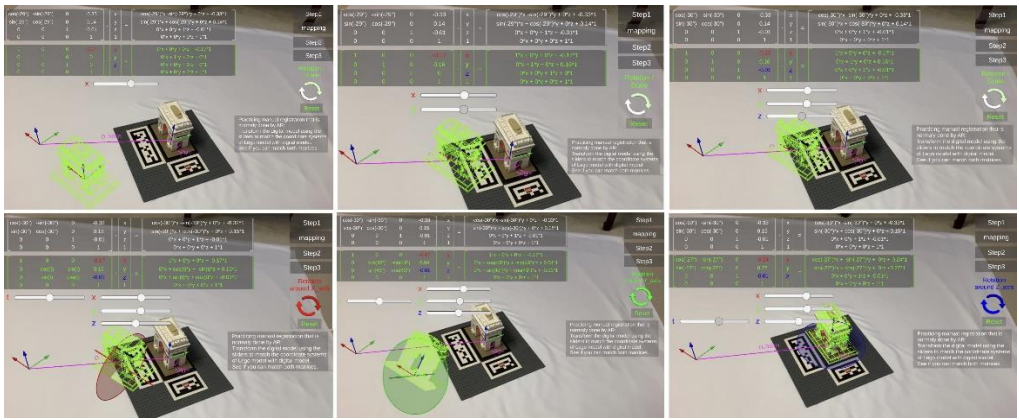

**Fig 9.** Practicing AR registration through playing with function parameters

## 4   Discussion:

The AR-enabled RTS prototypes intend to support architecture students learning parametric modeling, specifically geometric transformations and allied, essential mathematical concepts [3]. During the RTS tasks, spatial relations help students intuitively understand the math concepts behind geometric transformations. Some examples of the relations between spatial reasoning and math in the RTS tasks are described as follows:
- The spatial rotation in different directions (clockwise or counterclockwise) is associated with and displayed through the - and + signs in math. The arc graphically visualizes the rotation quantity with the corresponding degree notation.

- The measuring line graphically demonstrates the distance between the images of the transformations (pre-image and image). The distance notation also numerically matches with the physical distance. For example, when the distance is doubled, the measuring line and the distance notation demonstrate a doubled value.
- The transformation matrices numerically match with the graphical information of the transformation dynamically displayed through lines and arcs. The linear algebraic equations corresponding to matrix multiplication also numerically match with the graphical information of the corresponding transformations.

We expect that providing a spatial context for mathematical information associated with geometric transformations will positively help students in spatial cognition associated with geometric transformations and mathematics. We hope that such spatial contextualization help students understand math subjects such as matrices as an integrated procedure with spatial motions rather than solely numbers and equations. This positive effect may be most pronounced in students with medium and low spatial skills because this intervention may help them in constructing visual imagery of transformation matrices that could lead to understanding the math concepts more effectively in an intuitive way. Ultimately, we anticipate that this intervention may boost students' understanding of parametric modeling methods and use of the software tools accordingly.

## 5    Conclusion and Future work

In this paper, we have presented an AR educational prototype, named BRICKxAR_T, to improve learning fundamentals of parametric modeling. The goal of the BRICKxAR_T is to assist students in learning geometric transformations and their associated math functions as one of the essential components of parametric modeling. In this paper, we broke down the geometric transformation process into three levels of 'motion, mapping, and function' to help students understand the mathematical functions behind transformation actions. The AR educational prototype is designed to help students learn about digital modeling components named as variables, parameters, and functions through a transformation puzzle game. We believe that learning the fundamentals of parametric modeling and understanding the mathematical reasoning behind "digital design thinking" in an intuitive way may positively impact students in the professional use of parametric modeling methods for architectural design. We are going to evaluate the impact of BRICKxAR_T on students' spatial and math skill learning in future studies upon approved IRB (number: IRB2020-1213M). We are going to hold workshops conduct user studies to compare experimental and control groups, using AR and non-AR learning methods, respectively. We expect that contextualizing math functions in a spatial and physical scenario and visualizing the related information and notations may help students improve their understanding of geometric transformations as mappings and functions rather than motions. Also, we believe that BRICKxAR_T may reduce the mental load in learning geometric transformations and increase students' motivation in learning the targeted subject. In future studies, we are going to conduct multiple tests and surveys, including Purdue Visualization of Rotations Test [36], math tests on transformation matrices, NASA_TLX survey [37], and motivation surveys [38] to assess our claims.

## Acknowledgements

The research is funded by the Texas A&M University's Presidential Transformational Teaching Grant (PTTG), the Innovation [X] grant, and the Translational Investment Fund from the Innovation Partners.